\documentclass[10pt,a4paper]{article}
\usepackage{geometry}

\usepackage[utf8]{inputenc}

\usepackage{cite}

\usepackage{balance}       
\usepackage{graphics}      
\usepackage[T1]{fontenc}   
\usepackage{txfonts}
\usepackage{mathptmx}
\usepackage[pdflang={en-US},pdftex]{hyperref}
\usepackage{color}
\usepackage{tabularx}
\usepackage{multirow}
\usepackage{booktabs}
\usepackage{textcomp}
\usepackage{siunitx}
\usepackage{array}
\usepackage[gen]{eurosym}
\usepackage[dvipsnames, table]{xcolor}
\usepackage{cite} 

\usepackage{microtype}        
\usepackage{ccicons}          

\usepackage{todonotes}

\usepackage{xspace}

\usepackage{mdframed}
\definecolor{highlightboxColour}{rgb}{0.95,0.95,0.95}
\mdfdefinestyle{highlightboxmdf}{skipabove=5pt, skipbelow=10pt, linewidth=0pt, innertopmargin=5pt, innerbottommargin=5pt, backgroundcolor=highlightboxColour}

\usepackage{url}

\usepackage{microtype}
\DisableLigatures[f]{encoding = *, family = * }

\usepackage[aboveskip=1pt,labelfont=bf,labelsep=period,singlelinecheck=off]{caption}

\makeatletter
\renewcommand{\@biblabel}[1]{\quad#1.}
\makeatother

\usepackage{lastpage,fancyhdr,graphicx}
\usepackage{epstopdf}
\fancyheadoffset[L]{2.25in}
\fancyfootoffset[L]{2.25in}

\usepackage{color}

\definecolor{Gray}{gray}{.25}

\usepackage{graphicx}

\usepackage{sidecap}

\usepackage{wrapfig}
\usepackage[pscoord]{eso-pic}
\usepackage[fulladjust]{marginnote}
\reversemarginpar

\definecolor{rowcol}{rgb}{0.9,0.9,0.9}

\begin{document}
\vspace*{0.35in}

\begin{flushleft}
{\Large
\textbf{\newline{What is ``Intelligent'' in Intelligent User Interfaces?\\A Meta-Analysis of 25 Years of IUI}}
}
\newline
\\
Sarah Theres V\"olkel\textsuperscript{1*},
Christina Schneegass\textsuperscript{1},
Malin Eiband\textsuperscript{1},
Daniel Buschek\textsuperscript{2}

\bigskip
{1} LMU Munich
\\
{2} University of Bayreuth
\\
\bigskip
* sarah.voelkel@ifi.lmu.de

\end{flushleft}

\begin{abstract}
This reflection paper takes the 25th IUI conference milestone as an opportunity to analyse in detail the understanding of intelligence in the community: Despite the focus on intelligent UIs, it has remained elusive what exactly renders an interactive system or user interface ``intelligent'', also in the fields of HCI and AI at large. We follow a bottom-up approach to analyse the emergent meaning of intelligence in the IUI community:
In particular, we apply text analysis to extract all occurrences of ``intelligent'' in all IUI proceedings. We manually review these with regard to three main questions: 1) What is deemed intelligent? 2) How (else) is it characterised? and 3) What capabilities are attributed to an intelligent entity? We discuss the community's emerging implicit perspective on characteristics of intelligence in intelligent user interfaces and conclude with ideas for stating one's own understanding of intelligence more explicitly.

Author preprint, to appear in \textit{Proceedings of the 2020 Conference on Intelligent User Interfaces (IUI'20)}.
\end{abstract}

\section{Introduction}
\textit{Intelligent User Interfaces} are the eponymous subject of study at the IUI conference, which celebrates its 25th anniversary this year. 
Interestingly, it still remains difficult to define what exactly renders an interactive system or user interface ``intelligent'', not only at IUI but also in the intersecting fields of Human-Computer Interaction (HCI) and Artificial Intelligence (AI) at large. This shows both in a lack of widely-referenced or recent objective definitions and also our subjective experiences, for example, in workshop discussions with fellow IUI researchers on the topic.

If we have managed so far, should we bother with such terminological considerations? We argue that our understanding of intelligence in UIs can implicitly shape our work, for example, by influencing our assumptions, methods, and metrics, and not least simply by what we consider to be in-scope of IUI work. As highlighted in similar reflections~\cite{Hornbaek2017}, our understanding of such a core concept thus serves as a \textit{thinking tool} for our research. It seems adequate to examine such a tool from time to time to surface and make explicit its assumptions, usages, and boundaries.

Explicating our collective understandings of intelligence in UIs also serves as a \textit{communication tool}: Developing conceptual clarity contributes to a common belief system and terminology which is crucial for fostering productive discussions and exchanges, in particular for interdisciplinary research communities, such as IUI. 

The literature demonstrates that such reflections trigger important discussions: For example, Kostakos~\cite{Kostakos2015} highlighted the ``big hole'' in HCI research, based on a co-word analysis of 20 years of CHI conference proceedings~\cite{Liu2014}. The main critique therein is a lack of ``motor themes'', which are both central to a community and dense (internally coherent). In other words, HCI appears to lack core themes sufficiently developed to become mainstream -- as opposed to jumping between novel themes from year to year. CHI is a broad HCI conference. In contrast, IUI seems to directly carry such a motor theme in its name and focus -- intelligent user interfaces. Therefore, taking intelligent UIs as a \textit{central} theme in the IUI community, we examine whether it is also \textit{dense}, in the sense of being coherent in its understanding of the underlying core concept.

In particular, in this reflection paper, we take the 25-year milestone as an opportunity to analyse in detail the meaning of ``intelligent'' in the IUI community:
To do so, we applied text analysis to extract all occurrences of ``intelligent'' in all IUI proceedings. We then manually reviewed these with regard to three main questions: 1) What is deemed intelligent? 2) How (else) is it characterised? and 3) What capabilities are attributed to an intelligent entity?

Our results show a growing variety over the years of what is referred to as intelligent and how it is characterised as such. At the core, IUI researchers tend to describe intelligence in their work with the key aspects of automation, adaptation, and interaction. Researchers also associate different descriptors with intelligent technology depending on how it is conceptualised (e.g. as a UI, system, or agent). Moreover, we extract nine challenges that researchers mention in the context of intelligent entities.
\section{Background}

\subsection{Large-Scale Literature Text Analysis}

Recent work has brought forth examples of literature reviews based on large-scale text analysis:
For example, Liu et al.~\cite{Liu2014} analysed co-occurren\-ces of the ``author keywords'' sections in 20 years of CHI papers. Using clustering analysis, they derived core, popular, and backbone topics, and examined keyword networks. They discussed trends between two periods and the emerging structure of HCI research, revealing that it lacks ``motor themes'' that are both central and dense. 
Similarly, Lee et al.~\cite{lee2019weaving} analysed the citation network of CHI papers and investigated keywords in the abstracts to understand long-term topic trends. Their work identified emerging research topics and showed the rise and fall of different topics over the past three decades of CHI.
As mentioned in the introduction, these approaches motivate our analysis of a more specific venue with a stated focus, that is, intelligent user interfaces. In contrast to Liu et al. we analyse the papers' full text for a specific key term (``intelligen*'', i.e. intelligen-t/ce/tly etc.). We then also study co-occurrences of this key term with other descriptions (e.g. an entity might be called both ``intelligent'' and ``adaptive''). This allows us to gain insights into the researchers' implicit understanding of what is considered intelligent.

In another large-scale literature review, Abdul et al.~\cite{Abdul2018} recently used topic modelling to analyse over 12,000 papers from HCI and Machine Learning (ML)/AI regarding aspects such as explainability of intelligent interactive systems. Their resulting topic networks gave insights into the connections between research fields. As a key implication, the authors called for more interdisciplinary research at the intersection of HCI and AI to address the emerging challenges. Since IUI is placed at precisely this intersection, this motivates our detailed analysis of IUI proceedings in this paper. In particular, we focus on the emerging understanding of the core concept of intelligent systems and user interfaces in this interdisciplinary research.

Similar approaches have also been applied within the IUI community to analyse not the literature but the \textit{users}' views on interactive intelligent systems: Eiband et al.~\cite{Eiband2019} applied topic modelling to 35,448 reviews of apps with algorithmic decision-making on the Google Play Store. Combined with manual analysis, they revealed the problems that users faced during interaction. In contrast, our work focuses on the perspectives of \textit{researchers} on intelligent user interfaces and systems.

In another recent literature reflection, Hornb{\ae}k et al.~\cite{hornbaek2019we} analysed 35 years of CHI proceedings regarding the use and understanding of the term ``interaction''. They categorised 2,668 distinct terms used to modify and further describe interaction. Their results show growing diversity in such characterisations and that these styles of interaction play an increasing role in discourse in the field. In a similar manner, in this paper, we examine the use and description of the key term ``intelligen*'' in IUI proceedings.

\subsection{What is Intelligence?}
Intelligence is an intrinsic \textit{human} characteristic, and as such, has been transferred to computer science. This transfer shows, for example, in Turing's article from 1950, where he asked: ``Can machines think?''~\cite{turing1950computing}, and which can be considered the beginning of the discussion on the nature of technological intelligence. 
However, finding a common understanding and conceptualisation of \textit{human} intelligence in the first place has been the subject of intense scientific discussion in the past. To date, there is no single definition, which may partly account for the current diversity of notions in our field.

In the following, we first give a short overview of exemplary definitions of human intelligence as presented in psychology, before we compare them to example definitions of technological intelligence. 

\subsubsection{Human Intelligence}
In 1921, the discussion about human intelligence was the subject of the symposium of Educational Psychology on ``What do I conceive intelligence to be?''. A number of experts in the field shared their views on the question, arguing that, for example, \textit{thinking abstractly} or having the \textit{ability to learn and respond to questions} are essential for intelligence (cf.~\cite{lanz2000concept,Britannicaintelligence}).

In later years, the capability to \textit{adapt to the environment} has been introduced as another central part of intelligence. For example, Sternberg~\cite{sternberg1997concept} argues that ``intelligence comprises the mental abilities necessary for adaptation to [...] any environmental context [...] It offers people an opportunity to respond flexibly to challenging situations''~\cite{sternberg1997concept}. The capability of adaptation to new contexts is also emphasised by the American Psychology Association, describing intelligence as ``[...] the ability to derive information, learn from experience, adapt to the environment, understand, and correctly utilize thought and reason''~\cite{APAintelligence}. 

Besides trying to find general prerequisites for intelligence, the concept is often broken down into sub-types. For example, Sternberg defines intelligence via emotional, practical, analytical, and creative intelligence~\cite{sternberg1997concept}. Gardner even suggests that the brain consists of eight different forms of intelligence operating in autonomy, namely linguistic, logical-mathematical, spatial, musical, physical-kinaesthetic, interpersonal, and intrapersonal intelligence~\cite{gardner1983frames}. 

Within each individual human being, the manifestation of these different sub-types of intelligence can vary~\cite{gardner1983frames, gardner2003multiple}. This raised the question if intelligence should be seen as a holistic concept, or if it is rather certain capabilities that form intelligence (i.e., mathematical or linguistic skills).

\subsubsection{Technological Intelligence}
Especially early work involved in establishing IUI as a research area reflects on what it means for technology to be intelligent. On closer inspection, they take up different aspects from the definitions presented above.

For example, Stephanidis et al.~\cite{Stephanidis1997} in 1997 characterised IUIs ``[...] by their capability to adapt at run-time and make several communication decisions concerning `what', `when', `why' and `how' to communicate.'' Therein, they relate to an earlier view of a UI communicating concepts to the user~\cite{Szekely1991}.
They thus particularly highlight adaptation and decision-making, which in part resounds with Sternberg's~\cite{sternberg1997concept} definition and the one given by the American Psychology Association. Moreover, knowing ``what'', ``when'', ``why'' and ``how'' to communicate can be considered to involve some sort of interpersonal intelligence.

Similarly, in 2000, H\"o\"ok~\cite{Hook2000} reflected on ``Steps to take before intelligent user interfaces become real'': In her account, an important emerging view of IUIs was adaptation and personalisation.
Another related aspect is \textit{generating} UIs to automatically adapt to varying contexts, such as the device. This prominently appeared in 2004 in the \textit{SUPPLE} system by Gajos and Weld~\cite{Gajos2004}.

Moreover, Wahlster and Maybury~\cite{Wahlster1998} stated that intelligent user interfaces ``are human-machine interfaces that aim to improve the efficiency, effectiveness, and naturalness of human-machine interaction by representing, reasoning, and acting on models of the user, domain, task, discourse, and media (e.g. graphics, natural language, gesture).''
Their definition entails the ability to think and to understand language and gestures (linguistic and physical-kinaesthetic intelligence). Also, it defines the \textit{goal} or \textit{purpose} of technological intelligence -- efficiency, effectiveness, and naturalness of interaction.

These exemplary definitions give an idea of how technological intelligence is inspired by human intelligence or capabilities. Beyond that, it is shaped through the interaction context with users towards fulfilling a particular purpose.
Yet, these aspects often remain tacit in the literature.

With our work, we aim to shed light on \textit{what} (parts of) technology researchers attribute intelligence to in the IUI community (besides the eponymous term \textit{user interface}), how this intelligence is \textit{characterised}, and which \textit{capabilities} are ascribed to the intelligent entities.
We return to the listed definitions and reflections in our discussion at the end of this paper after presenting the results of our analysis.

\section{Method}
With our approach, we aim to gain deeper insight into the understanding of the term ``intelligen*'' as used in the IUI community. Thus, we analyse the utilisation of this notion throughout the past IUI conference proceedings. We follow the approach by Hornb{\ae}k et al. \cite{hornbaek2019we}, who extracted occurrences of the term ``interaction'' from the proceedings of 35 years of CHI. In our case, we extract occurrences of ``intelligen*'' from all IUI papers. Similar to their work, we then analyse these extractions and interpret themes and trends. 

\subsection{Data Acquisition and Extraction}
We obtained all papers published during the last 24 IUI conferences from 1993 until 2019 (note that there were no IUI conferences from 1994--1996) through a request to the ACM Digital Library. 
IUI as a conference has changed over two and a half decades and so have the  proceedings. Starting out as a small conference with 39 accepted papers in total in 1993, the IUI main proceedings comprised 71 full and short papers as well as adjunct proceedings including posters, tutorials, and workshops in 2019. For our analysis, we only included full and short peer-reviewed papers. To compile this set, we manually went through all IUI proceedings in the ACM digital library\footnote{\url{https://dl.acm.org/conference/iui}, accessed 17.01.2020}. This resulted in a data set of 1,111 papers. Please note that since we included both long and short papers, the number of accepted papers does not correspond with ACM's acceptance rates, which often only refer to long papers.
For these papers, we then removed all meta information\footnote{In particular, we removed title, author, affiliations, conference name, theme, data, location, session title, keywords, references, copyright statement.} so that only the text body was considered for our analysis.
We automatically extracted all sentences from the data set in which the term \textit{intelligent} or its grammatical derivatives (\textit{intelligence}, \textit{intelligential}, \textit{intelligently}) occurred at least once. 
We thereby acquired 1,804 sentences. 

\subsection{Data Analysis}
Due to the number and extent of CHI full texts, Hornb{\ae}k et al.~\cite{hornbaek2019we} relied on sentence parsing, which is limited in terms of the semantic detail that can be automatically captured. In contrast, the IUI dataset is small enough to allow us to combine a large-scale automated text analysis with an in-depth manual content assessment. Therefore, we could not only infer syntactical information on the usage of ``intelligent'' but also semantic information extracted from the context of the sentence.

In particular, our analysis focuses on three central aspects of the sentences:
\begin{enumerate}
    \item What is considered ``intelligent''? [\textit{entity}]
    \item Which other descriptors are used to further specify this intelligent entity? [\textit{co-descriptor}]
    \item What actions / capabilities are attributed to this intelligent entity? [\textit{action}]
\end{enumerate}

At first, all authors reviewed the 2,011 sentences to develop a code book for \textit{entities}, \textit{co-descriptors}, and \textit{actions}. We then selected a random sample of four sentences from each year's proceedings (resulting in 96 sentences), which were coded by four coders independently using the predefined coding rules. 
We compared the results and discussed all inconsistencies until we reached a consensus. Afterwards, we split the dataset among four coders for the final coding. 
In particular, we established the following coding procedures: 

\textbf{Entity:} We define the intelligent entity as any object, which is ascribed intelligence in the respective sentence. Hence, entities usually comprise one noun or a series of nouns, such as ``user interface'', ``system'', or ``information collection system''. For our analysis, we identified and labelled the entity based on this noun / series of nouns. After checking the use of \textit{interface} in the coded sentences, we merged \textit{user interface} and \textit{interface} into one entity.

\textbf{Co-descriptors:} Co-descriptors either refer to the intelligent entity (e.g. ``user-adaptive interactive systems''~\cite{Denaux2014})
or the described action (e.g. ``automatically extract human gestures''~\cite{tanveer2016automanner}) to further specify it. Here, we coded the infinite form of the adjectives as co-descriptors (e.g. \textit{user-adaptive}, \textit{interactive} and \textit{automatic} for the examples above). In a few cases, we merged multiple variations into one co-descriptor (e.g. \textit{automated}, \textit{automatic}, \textit{autonomous}; and \textit{context-sensitive}, \textit{context-dependent}, \textit{context-aware}). 

\textbf{Action:} The action describes what the intelligent entity does, i.e. its capabilities. We first manually identified the part of each sentence that refers to the action of the intelligent entity. These sentence parts could not be meaningfully counted when stated directly as they appear in the text (i.e. almost all occurred only once). Hence, we clustered them along their key verb (e.g. ``support [...] information seeking'' \cite{sun2006towards} to \textit{support}). Then, four coders discussed and merged clusters of verbs based on synonyms in this context (e.g. \textit{aid} and \textit{help}, \textit{collect} and \textit{gather}) and, similar to the process applied to entities and co-descriptors, variations of British English and American English (e.g. \textit{analyse} and \textit{analyze}).
We thereby obtained 60 distinct actions. Finally, we manually checked all resulting actions and corresponding codes and excluded single codes which did not fit the action verb cluster. Similarly, we went through the list of remaining codes to assign them to one of the key action verbs if possible.

\section{Results}

\subsection{Dataset Overview}
In total, we analysed 1,804 sentences containing our search terms. These come from 504 papers (out of 1,111 accepted papers) over 24 IUI proceedings (1993 - 2019). Each year on average contributed 75.17 sentences ($SD= 51.44$) from 21.00 papers ($SD=8.96$). Contributing papers on average contributed 3.58 sentences ($SD=4.86$). 

\begin{figure}[t]
  \centering
  \includegraphics[width=0.6\columnwidth]{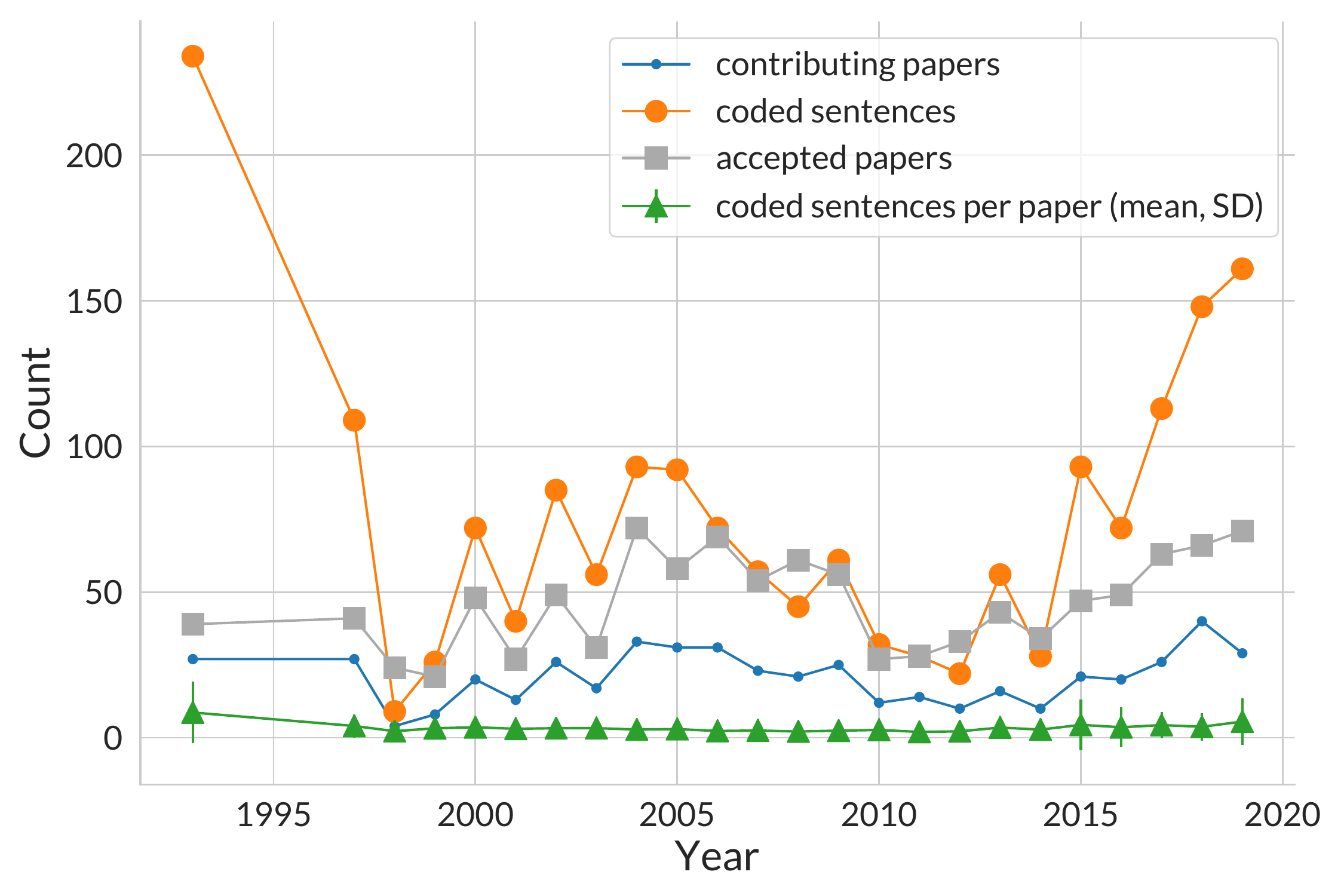}
  \caption{Overview of the coded data over time, in terms of absolute counts of coded sentences and papers (i.e. papers which include ``intelligen*'' at least once). The figure also shows the total number of accepted papers at each conference.}~\label{fig:coding_overview_over_time}
\end{figure}

Table~\ref{tab:coding_overview} shows summary statistics for our main coding categories (entities, co-descriptors, actions). Figure~\ref{fig:coding_overview_over_time} visualises the number of coded sentences and papers over time. The figure also includes the total number of accepted papers per year\footnote{\url{https://sigchi.org/conferences/conference-history/iui/}, accessed 17.01.2020}, showing that not every paper uses ``intelligen*'' (i.e. compare accepted papers and contributing papers).

\begin{table}[!t]
\centering
\small
\setlength{\tabcolsep}{4pt}
\begin{tabularx}{0.7\columnwidth}{Xrrr}
\toprule
& \textbf{entities} & \textbf{co-descriptors} & \textbf{actions} \\ 
\midrule
codings  & 2034 & 673 & 1308 \\
\quad mean (SD) per paper  & 4.22 (5.81) & 2.54 (2.48) & 3.79 (3.71) \medskip\\
distinct codes  & 522 & 250 & 59 \\
\quad mean (SD) per paper  & 2.50 (2.14) & 2.03 (1.53) & 2.99 (2.37) \medskip\\
papers (\% of all)  & 483 (95.83\%) & 266 (52.78\%) & 346 (68.65\%) \\
\bottomrule
\end{tabularx}
\caption{Overview of our coding. From top to bottom, rows show: number of codings per category (and mean / SD per analysed paper), number of distinct codes per category (and mean / SD), and number of papers that contributed at least one coding (percent of all contributing papers).}
\label{tab:coding_overview}
\end{table}

\subsection{Top Entities, Co-Descriptors, and Actions}

Figure~\ref{fig:coding_top_n} shows the histograms of the top codes per category. 

\subsubsection{Entities}
Our analysis yielded 2,034 codings of 522 distinct entities. 95.83\% of papers including the term ``intelligen*'' mentioned at least one intelligent entity with a mean of 2.50 ($SD=2.14$) distinct entities per paper (cf. Table~\ref{tab:coding_overview}). 

On the one hand, the entities which researchers labelled as \textit{intelligent} comprise different kinds of technology (e.g. ``system'', ``tool'', ``agent'') or components thereof (``interface'', ``technique'', ``algorithm''). On the other hand, they comprise a variety of functionalities or features (``assistance'', ``selection'', ``support''). 
The identified entities cover both intelligence in the interface versus intelligence internally in the system, as also discussed by Norman in 2002~\cite{norman2002complexity}.

The majority of coded entities (81\%) were only mentioned once due to their specificity, such as ``search support function''~\cite{Mauro:2018} or
``ToDo manager''~\cite{Shen2009}. On the other end, five entities were reported in twenty or more papers.

\begin{figure*}[t]
  \centering
  \includegraphics[width=\textwidth]{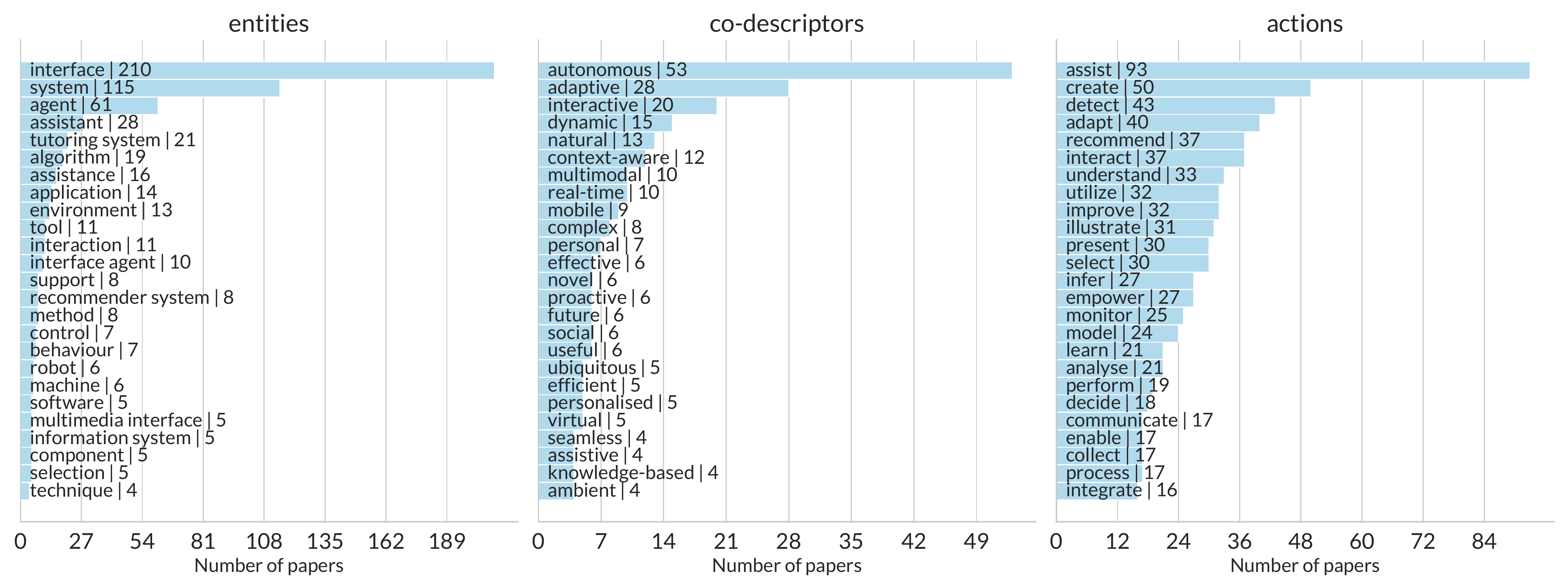}
  \caption{Overview of the top 25 occurring codes, ranked by number of papers in which they occur over all years.}~\label{fig:coding_top_n}
\end{figure*}

The eponymous \textit{intelligent user interface} is the most frequently mentioned entity, which is included in 210 papers (cf. Figure~\ref{fig:coding_top_n}). Another top entity is the \textit{intelligent system} (115 papers). Moreover, the entities \textit{agent} and \textit{assistant} are mentioned in 61 and 28 papers, respectively. Other potentially expected intelligent entities such as \textit{recommender system} and \textit{robot} were only directly described as intelligent in very few papers. 

Looking at the remaining entities beyond the top list in Figure~\ref{fig:coding_top_n}, we found, for example, further 62 distinct ``variations'' of \textit{system} with a total paper occurrence count of 76: For example, papers mention \textit{expert system} (3 papers), \textit{help system} (3), \textit{interface system} (3), and \textit{dialogue system} (3). All others were mentioned in fewer than three papers. 
Similarly, we found further 39 distinct ``variations'' of \textit{interface} with a total paper occurrence count of 43: These tend to refer to specific application contexts such as \textit{acquisition interface} (2 papers), \textit{visualization interface} (2), or \textit{cinematography interface} (1).
As another example of such specific applications, \textit{recommender systems} had ten further ``variations'' with a total paper occurrence count of eleven, including \textit{product recommender} and \textit{presentation recommender}.  

\subsubsection{Co-Descriptors}
Researchers used a variety of co-descriptors to further describe characteristics of intelligent entities. We found 250 distinct codes of these co-descriptors, which occurred 673 times in total. 52.78\% of the papers contributed to this list of co-descriptors with a mean of 2.03 distinct codings per paper ($SD=1.53$). 

Again, the majority of co-descriptors occurred only occasionally with eight co-descriptors mentioned in ten or more papers. 
The most frequently used co-descriptor is \textit{autonomous} (53 papers). For example, in 1993, Woods characterised an intelligent interface as ``an autonomous computer agent who mediates between process and practitioner''~\cite{woods1993}. 
The co-descriptor \textit{adaptive} was featured in 28 papers, for example, in 2004: ``The intelligence of the adaptive interfaces resides in the migration server, which adapts data collected at runtime [...]''
~\cite{bandelloni2004}. 
In 20 papers, intelligent entities were also labelled \textit{interactive}, for example, in 2013: ``These techniques raise major challenges for interactive and intelligent systems design to support the consumers' understanding and control of their energy usage in this complex environment.''~\cite{Fischer2013}. 
Another top co-descriptor is \textit{dynamic}. For example, in 2011 Bourke et al.~\cite{bourke2011} stated ``that the time is now right for more dynamic and intelligent interfaces to play an important role in a new generation of connected cameras''. 
Additionally, in thirteen papers designing \textit{natural} intelligent user interfaces was declared an important goal. For example, in 1997, Birnbaum et al. noted: ``Ironically, people often apply AI to interfaces to make them more natural, but without crafting the interface to support the intelligence, the AI component makes the interface far less usable''~\cite{birnbaum1997compelling}. 

\subsubsection{Actions}
In 68.65\% of papers, researchers described actions that intelligent entities perform. We coded 1,308 occurrences of such action descriptions, which we grouped into 60 key actions as described above. 187 codes could not be assigned to any cluster but describe individual cases, which are excluded for the following analysis. On average, we found a mean of 2.99 distinct action codes per paper ($SD=2.37$).

In 93 papers, an intelligent entity was used to \textit{assist}, aid or help users, e.g. in ``performing their tasks''~\cite{bixler2013detecting}, ``understanding and navigating information''~\cite{zhou2002semantic} or improving a variety of skills, such as for example ``develop writing proficiency''~\cite{bixler2013detecting} or improve ``public speaking''~\cite{tanveer2015rhema}. Moreover, the intelligence can support the user by guiding, facilitating, or easing a task, e.g. \textit{assisting} ``user orientation''~\cite{Rich:1997}, ``comprehending an information space''~\cite{Athukorala2016},
or ``understanding why a system decision has been made''~\cite{Eiband2019}.    
On a more functional level, intelligent entities \textit{create} or generate a huge variety of functions, content, or models, for example, ``simulations''~\cite{joyner2014}, ``explanations'' (e.g.~\cite{cai2019effects}), ``appropriate content for display''~\cite{harper2007talk}, etc.

Intelligent entities are also concerned with \textit{detecting} or capturing specific information, for example, to ``recognize the user's emotional state''~\cite{bosma2004exploiting} or to ``diagnose errors or misconceptions''~\cite{eisenberg1997helping}.

\begin{figure}[t]
  \centering
  \includegraphics[width=0.6\columnwidth]{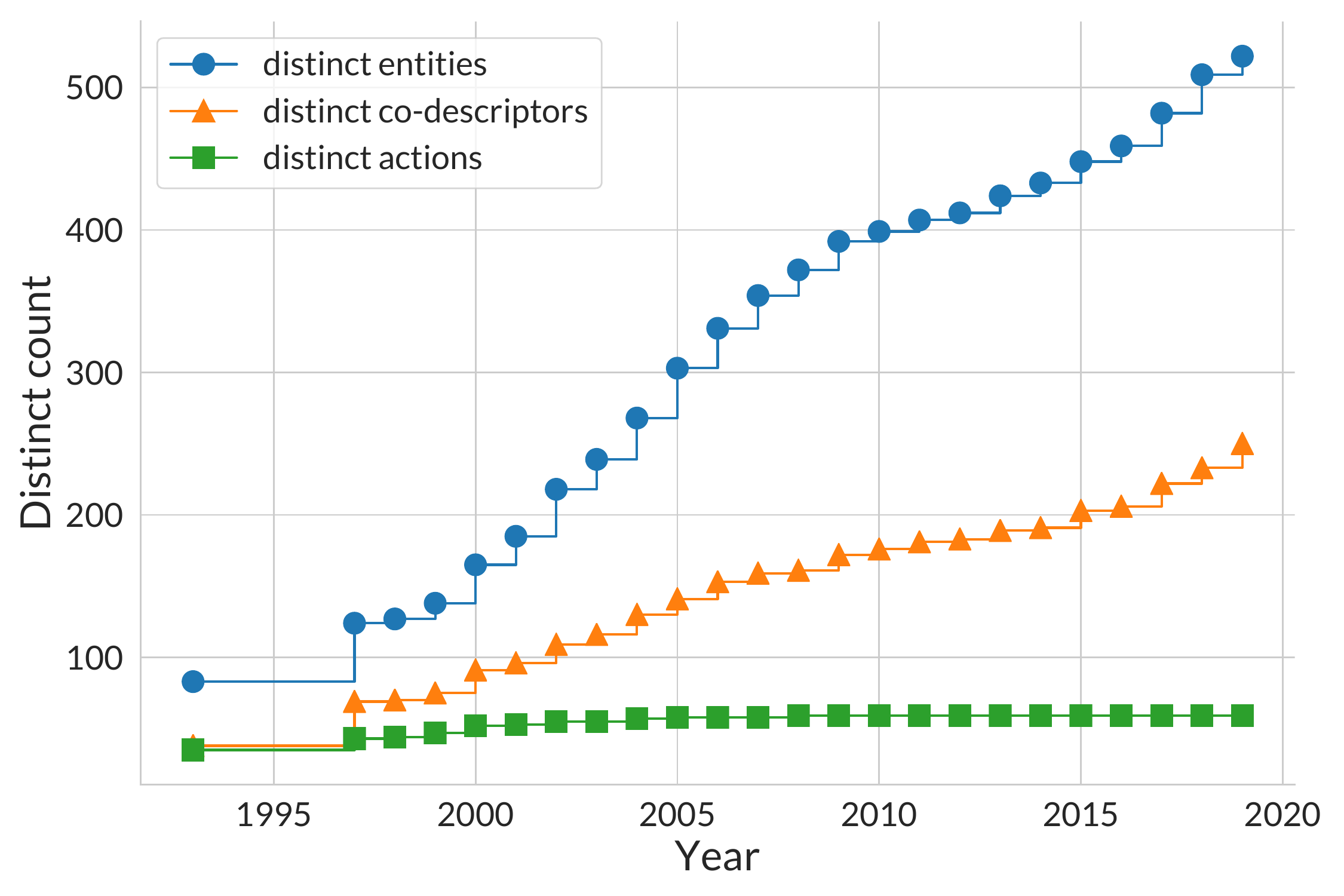}
  \caption{Development of total number of distinct entities, co-descriptors, and actions over time. Read: At each point, Y distinct codes have appeared until and including year X. This figure shows a growing diversity in what IUI researchers refer to as intelligent and how they describe it.  }~\label{fig:coding_distincts_over_time}
\end{figure}

In 37 papers, the task of intelligent entities was generally described as \textit{interacting} with the user (e.g. ``interact with learner''~\cite{qu2004choosing}) or parts of the entity (e.g. ``interacting intelligently with the natural language generation components''~\cite{lok2002ail}). Furthermore, several papers specified \textit{how} this intelligent interaction should be, for example, ``interact in a natural, more engaging way''~\cite{qu2004choosing}.

An important role of intelligent entities also seems to be to \textit{adapt}. On the one hand, papers describe to whom or what the intelligent entity \textit{adapts}, e.g. ``adapt to [the user] in real time''~\cite{li2017confiding} in general or specific user features (e.g. ``adapt [the agent's] behavior to that of the [user's] personality''~\cite{Xiao2019}). On the other hand, papers provide more information about which features of the intelligent entity are \textit{adapted}, e.g. ``adapt interface''~\cite{bandelloni2004} or ``adapt data in runtime''~\cite{bandelloni2004}. 

\subsection{Trends over Time}
We also analysed the occurrence of these entities, co-descriptors and actions over time. The main overall trend is an increasing variety of what is referred to as intelligent and how it is described, as visualised in Figure~\ref{fig:coding_distincts_over_time}: Each IUI conference added new entities and co-descriptors to the overall set, while the main actions were present already from very early on.

We found no clear trends regarding the numbers of (new) distinct counts per year.
We further examined the development of specific entities, co-descriptors, and actions over the years but also found no emerging trend patterns.

\subsection{Co-Occurring Codes}

\begin{figure}[t]
  \centering
  \begin{minipage}{0.45\textwidth}
        \centering
        \includegraphics[width=0.9\textwidth]{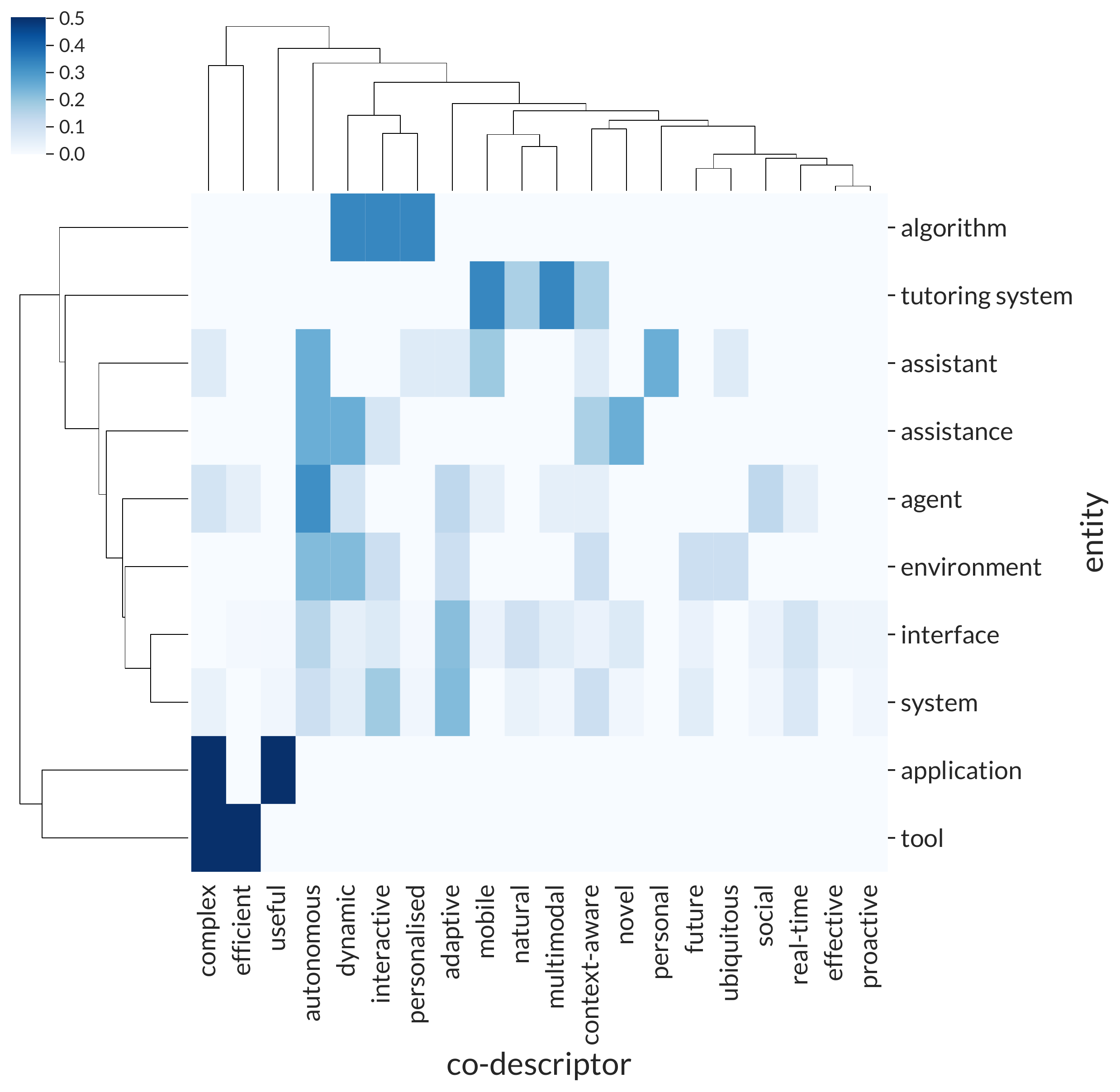}
        \caption{Co-occurrences of the top coded entities and co-descriptors. Colour shows the relative occurrence of co-descriptors per entity. The dendrogram shows the clustering.}~\label{fig:cooccurrence_entity_codescriptor}
    \end{minipage}\hfill
    \begin{minipage}{0.45\textwidth}
        \centering
        \includegraphics[width=0.9\textwidth]{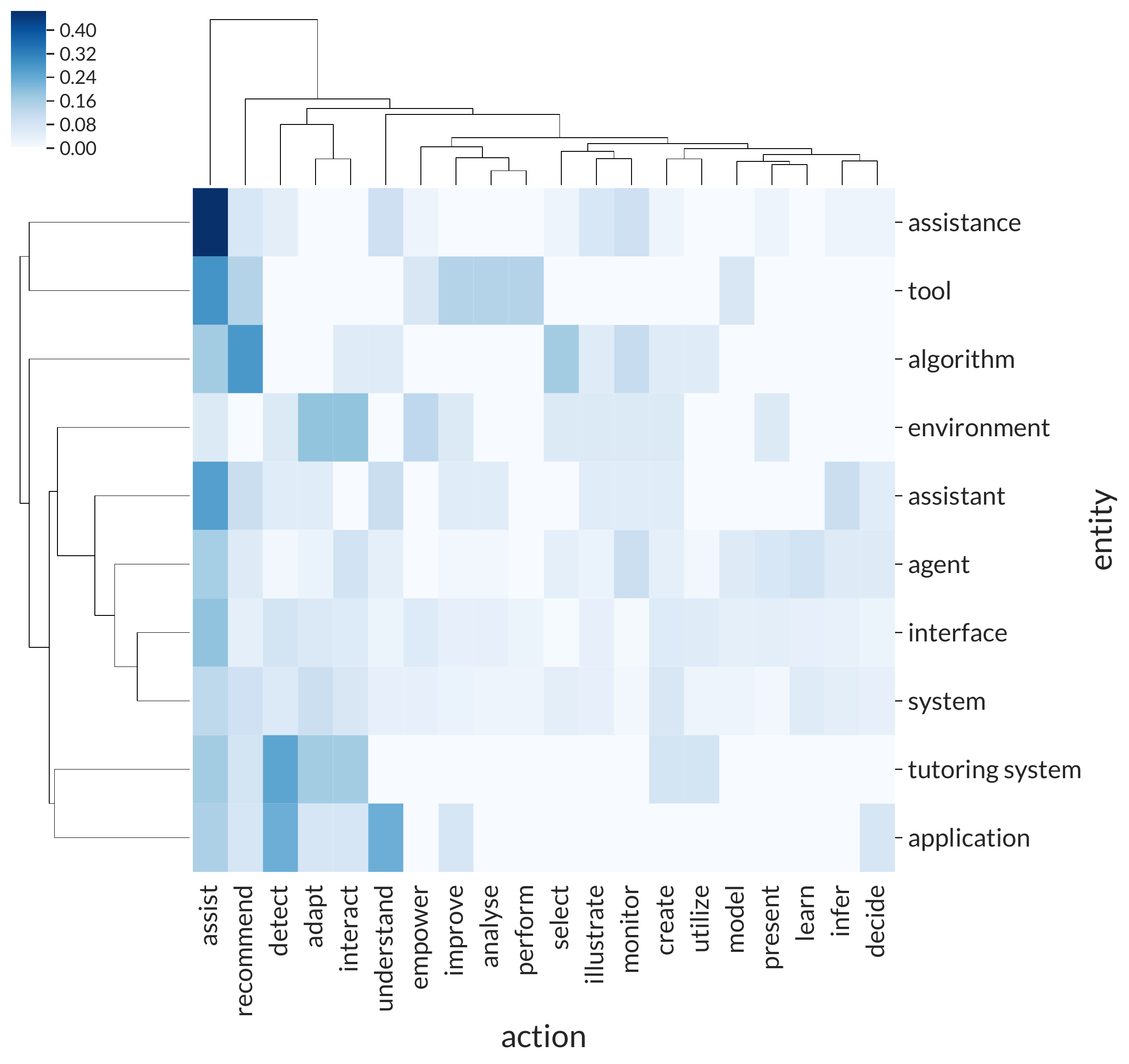} 
        \caption{Co-occurrences of the top coded entities and actions. Colour shows the relative occurrence of actions per entity. The dendrogram shows the clustering.}~\label{fig:cooccurrence_entity_action}
    \end{minipage}
\end{figure}

We further analysed how often different codes occur together: In particular, we counted 1) co-occurrences of entities and co-descriptors, and 2) co-occurrences of entities and actions.

For instance, the sentence \textit{``Evaluating the impact of expertise and route knowledge on task performance can guide the design of intelligent and adaptive navigation interfaces.''}~\cite{Ohn-Bar2018} contains both an entity (\textit{navigation interface}) and a co-descriptor (\textit{adaptive}).

For presentation and visual clarity, we applied clustering to arrange rows and columns of a co-occurrence clustermap. Informed by related work using clustering in literature analysis~\cite{Liu2014} we chose simple agglomerative (bottom-up) hierarchical clustering. Concretely, we used the \textit{seaborn}\footnote{\url{https://seaborn.pydata.org/generated/seaborn.clustermap.html}, accessed 17.01.2020} library for Python with the UPGMA algorithm~\cite{Sokal1958} and default settings. 

Figure~\ref{fig:cooccurrence_entity_codescriptor} shows such a clustermap for entities and co-descriptors. Note that the shown values are relative to each entity (i.e. co-occurrence counts in each row are divided by the row sum). This allows for a better visual comparison between entities.
Complementary, Figure~\ref{fig:cooccurrence_entity_action} shows the clustermap for entities and actions. Overall, these figures illustrate the variation between intelligent entities with regard to the co-descriptors and actions that researchers associate with them.

\subsection{Descriptions of Intelligence}
We also analysed descriptions of intelligence as a concept itself (i.e. ``intelligence'' as a noun). This occurs rather sparsely, 134 times in total. The top descriptors of intelligence in our data are \textit{artificial} (60 times), \textit{machine} (11), \textit{social} (10), \textit{ambient} (6), \textit{computational} (5), \textit{emotional} (5), and \textit{explainable} (5). Further 21 distinct descriptors of intelligence occurred less than five times each.

\subsection{Challenges in Designing Intelligent Technology}
Designing intelligent technology is challenging from a technical and a human-computer interaction perspective. We manually extracted such challenges as indicated in our data set through keywords like ``problem'', ``issue'', ``need'', and so on. In this way, we collected 117 occurrences of challenges which we clustered and summarised into nine overarching themes. We next describe these themes along with references to example papers. 

\subsubsection{Presentation in the Interface}
Several papers named the presentation of technological intelligence in the interface as a challenge. This refers to specific aspects, such as whether and how to present uncertainty~\cite{Fischer2013}, but also generally highlights the interface as a limited resource that needs to be managed carefully~\cite{ovans1993intelligent}.

\subsubsection{Negative Effects of Intelligence}
Our analysis revealed that researchers embrace the potential benefits of technological intelligence, but also caution against possible negative side effects with regard to users. Most often stated were user frustration due to a lack of algorithmic accuracy~\cite{vertanen2009, wong2011}, and potentially tedious interaction to influence the algorithm~\cite{mclean2011}. H\"{o}\"{o}k~\cite{Hook:1998} even argues that technological intelligence may hinder the development of clear mental models of the system and thus negatively impact interaction.

\subsubsection{Explanation}
A further challenge prevalent in our data set is the adequate explanation of system behaviour to support understanding of system suggestions, predictions, and decisions. From an HCI perspective, this entails assessing what users want and need to have explained~\cite{penney2018toward}. Mynatt and Tullio~\cite{mynatt2001inferring} highlight that fostering user understanding of technological intelligence becomes particularly important in wearable and ubiquitous computing.

\subsubsection{User Interaction}
Interaction with the user was repeatedly named as a crucial and multi-faceted challenge for designing intelligent technology. As Smith and Lieberman put it, ``people interact with interfaces to accomplish goals [...]''~\cite{Smith2010}. Along these lines, interaction requires establishing awareness of a shared context, reasoning about shared goals, planning and allocating resources to reach these goals, as well as communicating and evaluating progress~\cite{amant1998interaction}. 

This includes creating a suitable way to communicate about goals~\cite{amant1998interaction}, providing feedback options~\cite{Glowacka:2013}, and supporting error detection and correction ~\cite{miller1997computational, springer2019progressive}.
However, interaction needs ``balance between user control and autonomy [...] in a flexible way without overwhelming the user with requests, and without risking undesired system actions''~\cite{Fischer2013}.

\subsubsection{Technical Challenges}
The technical challenges involved in designing intelligent entities are manifold. With regard to training a learning system, difficulties referred to limited training data in early stages of use~\cite{wong2011} and finding patterns in the data~\cite{chi2011intelligent}. 
A closely related challenge is the management of false positives and recognition errors against the negative cost of reject rates~\cite{katsuragawa2017effect}. More generally, technological intelligence was linked to the control of complex environments~\cite{Fischer2013}, to the interaction via multiple channels or devices~\cite{fitzgerald2003multimodal, miller1997computational}, and to ensuring system robustness~\cite{miller1997computational}. Moreover, capturing user data such as their affective state~\cite{tan2013informing} or interpreting imprecise natural language instructions~\cite{lefort2017dimensions} was named as technically challenging. 

\subsubsection{When to Integrate Intelligence}
In the papers we analysed, researchers called for a considerate integration of technological intelligence. That is, a core challenge for designing intelligent technology is to reflect on \textit{when} to make technology intelligent in general~\cite{edmonds1993future}, to decide on \textit{where} to integrate intelligence~\cite{birnbaum1997compelling} and for what kind of tasks~\cite{Puerta1993}. 
These considerations need to balance the advantages of artificial intelligence techniques against those of traditional methods, and the system complexity against user benefits~\cite{birnbaum1997compelling}. This is in line with Nagel~\cite{nagel1998opportunity}, who argues that a challenge for designing IUIs is to prove that intelligent behaviour (e.g. adaptation) actually improves interaction.

\subsubsection{Trust, Confidence, Reliance and Privacy}
A central challenge prevalent in our data set was to establish user trust in the technology and confidence in its outputs (cf.~\cite{faulring2010agent, kristensson2005relaxing}), to demonstrate the trustworthiness of the technology~\cite{Fischer2013}, and to foster appropriate reliance~\cite{glass2008toward}. Moreover, privacy issues (e.g. through camera-based user tracking) were mentioned as a concern~\cite{avrahami2019unobtrusive}.

\subsubsection{Usability and Usefulness}
Usability, usability issues, and the violation of usability principles were named by many papers in our data set (e.g.~\cite{birnbaum1997compelling, glass2008toward, wong2011}) and decreased productivity and performance as a related concern~\cite{Kotowick2017,wong2011}. This also comprises the costs involved in verifying and correcting mistakes by intelligent technology, such as false inferences~\cite{birnbaum1997compelling}. Furthermore, in order to be useful, intelligent technology should consider users' changing tasks and needs~\cite{Bao2006}.

\subsubsection{Evaluation and Standards}
Finally, the analysed papers called for evaluation criteria (e.g. for visualisation techniques~\cite{dostal2013subtle}) 
and standards, for example for transparency~\cite{eiband2018bringing} and interface design~\cite{miller1997computational}.
\section{Discussion}

\subsection{Limitations}
Investigating the notion of ``intelligence'' in IUI through automated text analysis of how it is used in the written proceedings comes with a set of limitations:

The ACM Digital Library provided us with xml files for all IUI conference proceedings, which included the full text and meta information for each paper. These full texts were likely generated by automatically extracting text from pdfs. Hence, single information or texts were missing and parsing errors occurred (e.g. not recognising ``fi'' as two separate characters). We addressed this challenge by manually searching for missing parts. However, we cannot rule out that occasional occurrences of ``intelligen*'' could not be detected by our automatic analysis.

Furthermore, we extracted single sentences that include ``intelligen*''. Hence, the context of the paragraph is not considered, potentially missing further definitions or specifications of the term. For example, indirect mentions through grammatical placeholders or antecedents in subsequent sentences could reveal more information on the characteristics of ``intelligent'' entities (e.g. ``We implemented an intelligent system. \textit{It} is characterised by [...].''). The same applies to the identified challenges, which should not be seen as a comprehensive overview of all challenges discussed at IUI. Rather, we can think of them as those challenges typically mentioned when describing something as intelligent. 

Examining the entire paragraph rather than only the sentence for each occurrence of ``intelligen*'' could also give further insights into how deeply the notion of intelligence is considered in the individual papers. For example, the use of ``intelligen*'' could either be limited to a single sentence, e.g. to label a system or be embedded in a paragraph of deeper discussion.

Moreover, for the scope of this work, we did not include synonyms of intelligent, such as ``smart'', or abbreviations (e.g. ``AI'' -- Artificial Intelligence, or ``ITS'' -- Intelligent Tutoring System). 

We favoured greater semantic detail through manual analysis of the extracted sentences over further automated text analysis on a larger data set (e.g. including further proceedings such as CHI; cf. the interaction analysis of Hornb{\ae}k et al.~\cite{hornbaek2019we}). 
Yet, our analysis only covers explicit evidence of the use of ``intelligen*'' in IUI papers and assigning them to a set of themes. Thus, implicit meanings and conceptions of intelligence may not be depicted in these themes. For example, an understanding of intelligence could also be derived from the characteristics of systems that were presented in IUI papers. Since such an analysis would require an in-depth reading of all articles, it was out of scope for this paper.

\subsection{Adaptation, Automation, and Interaction are at the Core}

Our analysis shows that \textit{adaptation}, \textit{automation}, and \textit{interaction} are the most common aspects that IUI researchers highlight when describing something as intelligent. These co-descriptors are present across all years. Based on our analysis, we view them as the most coherent core of intelligence in IUI work. 

These findings align with early reflection on IUIs: For example, H\"o\"ok~\cite{Hook2000} also emphasised adaptation and personalisation in her account of IUIs in 2000. Similarly, Stephanidis et al.~\cite{Stephanidis1997} early on highlighted that IUIs adapt the way they ``communicate'' with the user.
Moreover, Wahlster's~\cite{Wahlster1998} definition of IUIs in 1998 (see Background Section) starts its list of descriptors for IUIs with improving efficiency and effectiveness, which also often seems to underlie the motivation for adaptation and automation.
Regarding their combination, both adaptation and automation are, for example, married prominently in one of the most widely referenced IUI papers: \textit{SUPPLE}~\cite{Gajos2004} was a system for automatically generating UIs to adapt to different devices and usage patterns.
Finally, interaction arguably underlies the HCI side of IUI in general throughout all such work.

While automating a process is also included in the most frequently mentioned actions (e.g. \textit{creating, detecting, recommending}), our list of actions underlines the importance of IUIs to primarily \textit{assist} the user. These findings indicate that in the discussion of the role of artificial intelligence to replace or augment humans, the IUI community emphasises the importance of assisting or empowering humans rather than replacing user skills. 

In conclusion, adaptation, automation, and interaction provide a common denominator for what many researchers regard as intelligent about IUIs. There are, of course, countless different ways that researchers envision this to unfold, for instance, concerning how automation is integrated into an interactive system.

\subsection{IUI Embraces both (Tool-) UIs and Agents}

A key aspect in H\"o\"ok's~\cite{Hook2000} early reflection about IUIs concerned tools and systems vs. agents: \textit{``Here we must be careful of imitating human-human communication and assuming that that would be the best model of performance.''} 
This echoes the longstanding wider discussion of whether HCI should embrace intelligent technology as agents or as tools. Most prominently, these views have been debated by Shneiderman and Maes (e.g. in 1997~\cite{Shneiderman1997}), and this is still ongoing and relevant twenty years later~\cite{Farooq2017}. 

In our analysis, we indeed see this fork in the path appear for the IUI community as well: One the one hand, researchers describe user interfaces, systems, and tools as intelligent. On the other hand, we see descriptions of intelligent agents and assistants.
This reveals that IUI has broadly embraced both intelligence for more traditional UIs (e.g. GUIs), as well as for agents, such as chatbots and voice agents.

Our analysis reveals diverging tendencies of what intelligence entails for such different entities: For example, intelligent interfaces and systems tend to be described relatively more as \textit{adaptive} and \textit{interactive}, compared to agents and assistants. These, in turn, tend to be described relatively more as \textit{autonomous}. Assistants are also often labelled as \textit{personal}. This suggests that researchers' implied assumptions and goals around intelligence vary depending on whether the technology is seen, for example, as a UI, system, tool, or agent.

\subsection{Examining Emerging and Explicit Descriptions of Intelligence}
Equipped with the terms of our analysis, we can also critically examine how these terms are related to definitions of intelligence: For example, taking a definition by Singh~\cite{Singh1994}, a system is intelligent if we need to ``attribute cognitive concepts such as intentions and beliefs to it in order to characterize, understand, analyze, or predict its behavior''. 

Therein, our most common descriptors seem to be reflected as follows: Adaptation can be connected well to changing behaviour, for example, based on changing ``beliefs'' of the system about the state of the world (e.g. the user's context for context-aware UIs). In addition, automation brings a certain independence from human actions, which could be seen as implying own intentions and beliefs (e.g. in voice assistants that perform tasks for the user). 

Moreover, comparing the identified actions of intelligent entities with the understanding of human intelligence (see Section 2.2.1), it is interesting to note that apart from the ability to think abstractly~\cite{lanz2000concept,Britannicaintelligence}, the other capabilities associated with human intelligence -- such as \textit{learning}~\cite{lanz2000concept,Britannicaintelligence}, \textit{responding to questions}, \textit{adapting to the environment}~\cite{sternberg1997concept}, \textit{understanding}~\cite{lanz2000concept,Britannicaintelligence} -- are among the most frequently mentioned actions. 

In this way, the revealed descriptors and actions can help us to examine previous and future definitions of intelligent technology in the light of the emerging views from the literature (and vice versa).

\subsection{Diversification}

Our results show a clear trend towards varying and diversifying descriptions of intelligence (Figure~\ref{fig:coding_distincts_over_time}): Each IUI conference published papers that referred to new entities as intelligent, accompanied by new co-descriptors. Such new references to intelligence might arise due to changing trend terms, new emerging technology and concepts, and more fine-grained specifications of previous ones (e.g. system vs. dialogue system). Overall, this matches the results of the analysis of Hornb{\ae}k et al.~\cite{hornbaek2019we}, who found diversifying characterisations for the concept of ``interaction'' in CHI proceedings. 
At the same time, we observed that most of the fundamental actions in our coding were present already early on. Together, these results thus paint the picture of a growing diversity of intelligent technology being used for fundamental actions in an interactive context, most prominently assisting and supporting users in their tasks.

\subsection{Challenges}

We collected nine challenges that emerged from the coded sentences. Together, they show the differentiated perspective of our community on intelligent technology: Technological intelligence is not presented as unconditionally beneficial but should rather be considerately integrated, weighing advantages and drawbacks with regard to user interaction. For example, although intelligent technology may help users to do certain tasks more efficiently, this support may come with usability issues and increase the overall complexity of interaction. 

One of the identified challenges points to missing standards, both in terms of evaluation and interface design for intelligent technology. To date, the community often still seems to lack such standards and best practices, and future research might work towards finding ways to establish them.

\subsection{Where Do We Go from Here?}
Our work provides a survey of the status quo of the meaning of ``intelligent'' based on 25 years of active research in the IUI community.
Interestingly, we found very few cases where researchers attempted to \textit{explicitly} define how they understand intelligence in their work. Rather, our community currently assumes -- or relies on -- an implicit understanding of ``intelligent'', which is also evident from the considerable number of IUI papers that never refer to ``intelligen*'' at all.

As one step towards more explicit engagement with the key terminology of our research subject, our analysis suggests that we can explicitly state how our assumptions and goals realise -- or relate to -- \textit{adaptation}, \textit{automation}, and \textit{interaction}. These emerged as the three key aspects covered by our community's understanding of ``intelligent''. Moreover, explicitly stating, for example, where we position our work in the \textit{tools vs agents} discussion can help us and the readers of our papers to contextualise our assumptions and goals around the involved intelligence.

That being said, we expect that insights and influences from HCI, AI, and related fields will continue to shape our understanding of ``intelligent'', such as intelligence as the ability to learn (c.f.~\cite{chollet2019}).
We therefore should continue the discussion of our understanding of the concept of intelligence and its key aspects beyond this paper.

\section{Conclusion}

More and more technology seems to be labelled as ``intelligent'' yet it has remained elusive what exactly renders an interactive system or user interface deserving of this term. 

In this reflection paper, we analysed in detail the understanding of intelligence in the IUI community with its focus on research at the intersection of HCI and AI. 
In particular, we followed a bottom-up approach to analyse the emerging implicit perspective on characteristics of intelligence in intelligent user interfaces. To do so, we combined text scraping with manual analysis to examine 2011 sentences with ``intelligen*'' in the past 24 years of IUI proceedings. 

Our analysis revealed that IUI researchers tend to describe intelligence in their work with the key aspects of automation, adaptation, and interaction.
Moreover, researchers associate different descriptors with intelligent technology depending on how it is conceptualised (e.g. as a UI, system or agent).

These insights can help us as researchers at the intersection of HCI and AI to reflect on and more explicitly communicate our assumptions and goals around intelligence in our work.
For example, we could explicitly label and describe aspects of adaptation and automation in our work, or make an explicit decision for presenting intelligence as part of a tool, UI, agent, assistant, and so on.

Overall, this paper thus for the first time provides a detailed reflection on the emerging collective understanding of intelligence as a key concept in this interdisciplinary community.

\section{Acknowledgements}
We thank ACM and the ACM Digital Library for providing us with full texts for all IUI proceedings.
This project is funded by the Bavarian State Ministry of Science and the Arts in the framework of the Centre Digitisation.Bavaria (ZD.B).

\bibliography{bibliography}

\bibliographystyle{abbrv}

\end{document}